\documentclass[aps,prb,twocolumn,groupedaddress,showpacs,showkeys,amsmath,amssymb]{revtex4}

\usepackage{amsfonts}
\usepackage{amssymb,amsmath}
\usepackage{graphicx}
\usepackage{dcolumn}

\begin{document}

\title{Finite low-temperature entropy 
       of some strongly frustrated quantum spin lattices\\
       in the vicinity of the 	saturation field}

\author{Oleg Derzhko$^1$\footnote
        {On leave of absence from
        the Institute for Condensed Matter Physics,
        National Academy of Sciences of Ukraine,
        1 Svientsitskii Street, L'viv-11, 79011, Ukraine}
        and
        Johannes Richter$^1$\footnote
        {On leave of absence from
        Institut f\"{u}r Theoretische Physik,
        Universit\"{a}t Magdeburg,
        P.O. Box 4120, D-39016 Magdeburg, Germany}}
        
\affiliation{$^1$Max-Planck-Institut f\"{u}r Physik komplexer Systeme,
             N\"{o}thnitzer Stra\ss e 38, 01187 Dresden, Germany}

\date{\today}



\begin{abstract}
For a class of highly frustrated antiferromagnetic quantum spin lattices
the ground state exhibits a huge degeneracy in high magnetic fields 
due to
the existence of localized magnon states.
For some of these spin lattices (in particular,
the 1D dimer-plaquette, sawtooth and kagom\'{e}-like chains
as well as the 2D kagom\'{e} lattice)
we calculate rigorously 
the ground-state entropy at the saturation field. 
We find that
the ground-state entropy per site remains finite at saturation.
This residual ground-state entropy produces a maximum 
in the field dependence of the isothermal entropy at low temperatures.
By numerical calculation 
of the field dependence of the low-temperature entropy
for the sawtooth chain we find
that 
the enhancement of isothermal entropy  
is robust against 
small deviations in exchange constants. Moreover,
the effect is most pronounced in the extreme quantum case of 
spin $\frac{1}{2}$.
\end{abstract}

\maketitle

Antiferromagnetically interacting quantum spin systems
on geometrically frustrated lattices
have attracted much attention during the last years \cite{01,02,03}.
Whereas in general 
frustration makes the eigenstates of the quantum spin system very complicated, 
it has been found recently that in the vicinity of the saturation field
for a wide class of frustrated spin lattices 
just owing to frustration
the ground states become quite simple.
These exact ground states consist of independent localized magnons
in a ferromagnetic environment \cite{04,05,03}.
They lead to a macroscopic jump
in the zero-temperature magnetisation curve just below saturation \cite{04,05,03}
and may provide instabilities towards lattice deformations \cite{06}.

In the present paper
we examine the low-temperature entropy 
of several highly frustrated antiferromagnetic spin lattices
which may host independent localized magnons
in the vicinity of the saturation field.
The ground state of such a system at saturation 
exhibits a huge degeneracy 
which grows exponentially with system size.
For some of the considered spin systems
the ground-state degeneracy at saturation and therefore entropy 
can be calculated exactly by mapping the localized magnon problem 
onto a related  lattice gas model of hard-core objects.  
The latter models have been studied in many papers 
over the last few decades 
(see Refs. \onlinecite{07,08,09,10} and references therein).
We complete these analytical findings for the ground-state entropy 
by exact diagonalisation data 
for the sawtooth chain of $N=8,\;12,\;16$ sites
to extend our conclusions 
to fields below the saturation 
and 
to nonzero temperatures.
We also examine the effects of exchange anisotropy, 
different spin values $s$
and deviations from the condition on bond strengths
under which the independent localized magnons 
are exact eigenstates for the sawtooth chain. 
Finally, 
we discuss briefly the possibility 
of experimental verification of our findings.

We mention recent papers 
of Moessner and Sondhi \cite{11b}, Zhitomirsky \cite{11a} and  
Udagawa et al. \cite{11}
having some relation to our investigations. 
These authors 
calculate the ground-state degeneracy
of the  2D kagom\'{e} lattice carrying  {\it classical}
spins for certain spin configurations 
(up-up-down structure \cite{11b,11a} 
and structures 
obeying a ``modified ice rule'' \cite{11}) 
by mapping 
the spin problem onto a dimer-covering problem on the honeycomb lattice.
Udagawa et al. use their result to 
explain the residual entropy of the kagom\'{e} ice state 
which occurs in the spin ice compound Dy$_2$Ti$_2$O$_7$ 
under a magnetic field \cite{12,13}.
Note, however, that our study 
refers to the  frustrated {\it quantum}
spin lattices.
 
\begin{figure}[t]
\vspace{0cm}
\epsfxsize=7.5cm
\centerline{\epsffile{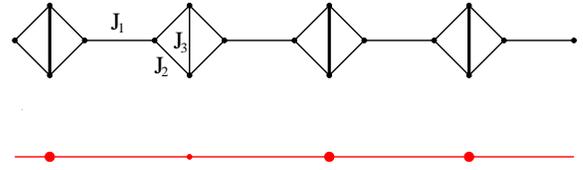}}
\caption
{The dimer-plaquette chain which hosts three localized magnons
at fat bonds 
(top)  
and the auxiliary lattice used for the calculation of 
the ground-state degeneracy 
at saturation 
(bottom). 
The localized magnons are eigenstates 
for large enough vertical bonds $J_3 \ge J_3^c(J_1,J_2)$
\cite{14}. 
\label{fig01}}
\end{figure}

\begin{figure}[t]
\vspace{0cm}
\epsfxsize=7.5cm
\centerline{\epsffile{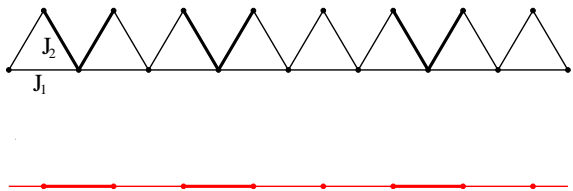}}
\caption
{The sawtooth chain which hosts three localized magnons
at fat {\sf{V}} parts
(top)
and the auxiliary lattice used for the calculation of 
the ground-state degeneracy at saturation
(bottom).
The localized magnons are eigenstates 
for $J_2 = \sqrt{2\left(1+\Delta\right)} J_1$ \cite{04,05}. 
\label{fig02}}
\end{figure}

\begin{figure}[t]
\vspace{0cm}
\epsfxsize=7.5cm
\centerline{\epsffile{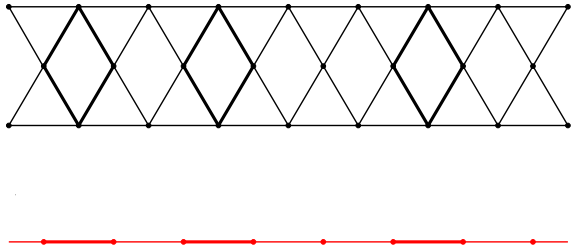}}
\caption
{The kagom\'{e}-like chain of Ref. \onlinecite{17} which hosts three 
localized magnons 
(marked by bold diamonds)
and the auxiliary lattice used for the calculation of 
the ground-state degeneracy 
at saturation. The localized magnons are eigenstates 
for exchange bonds of uniform strength  \cite{04}. 
\label{fig03}}
\end{figure}

\begin{figure}[t]
\vspace{0cm}
\epsfxsize=7.5cm
\centerline{\epsffile{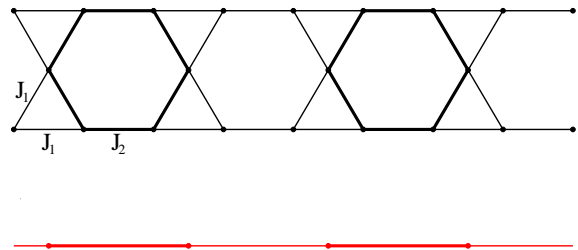}}
\caption
{The kagom\'{e}-like chain of Refs. \onlinecite{18,17} which hosts 
two localized magnons 
(marked by bold hexagons) 
and the auxiliary lattice used for the calculation of 
the ground-state 
degeneracy at saturation.
The localized magnons are eigenstates 
for $J_2 = \frac{1+2\Delta}{1+\Delta}J_1$ \cite{04}. 
\label{fig04}}
\end{figure}

\begin{figure}[t]
\vspace{0cm}
\epsfxsize=7.5cm
\centerline{\epsffile{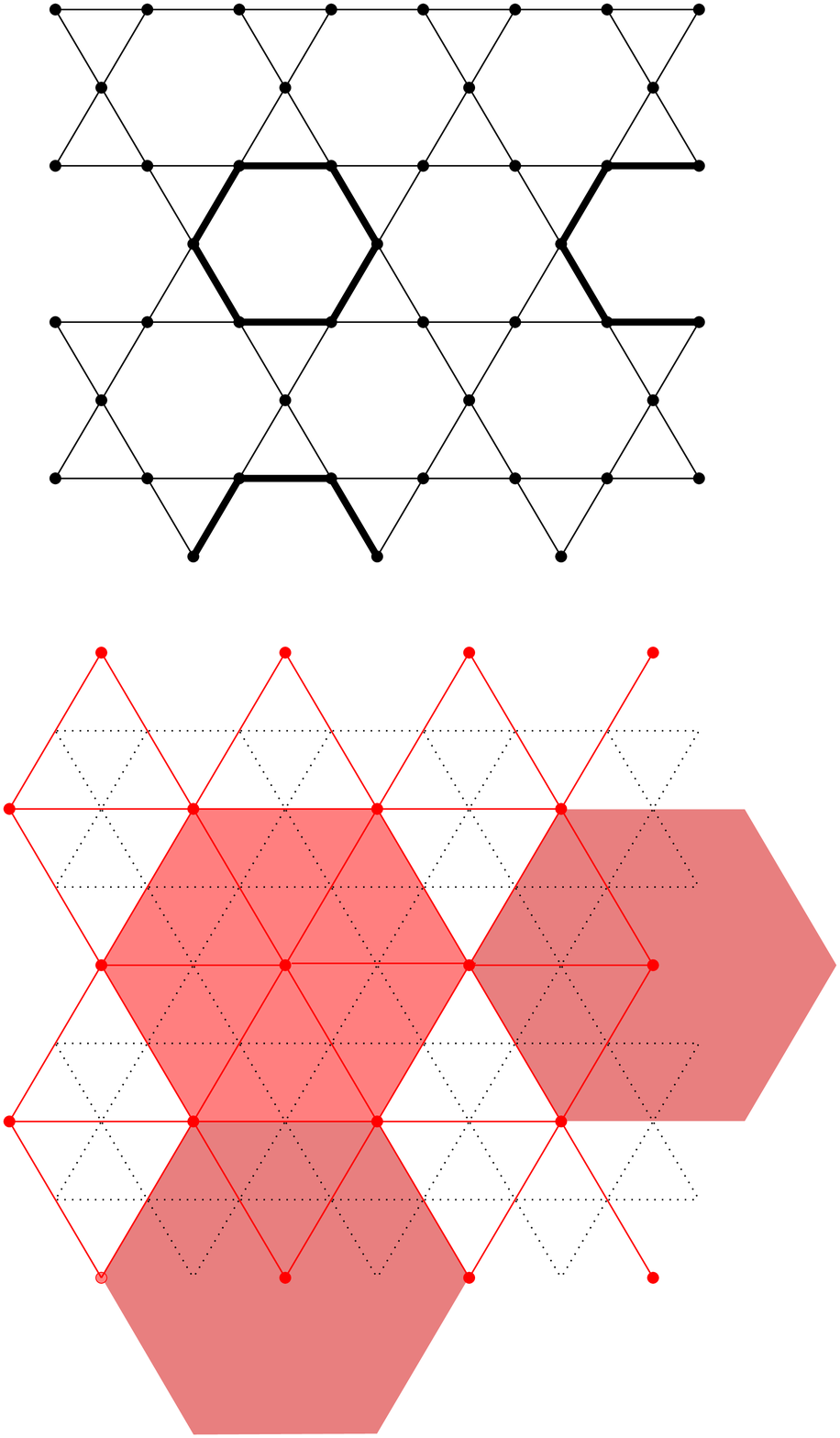}}
\caption
{The kagom\'{e} lattice which hosts three localized magnons 
(bold hexagons)
and the auxiliary triangular lattice with hard hexagons 
used for the calculation of the ground-state degeneracy at saturation.
The localized magnons are eigenstates 
for exchange bonds of uniform strength  \cite{04}. 
\label{fig05}}
\end{figure}

\begin{figure}[t]
\vspace{0cm}
\epsfxsize=7.5cm
\centerline{\epsffile{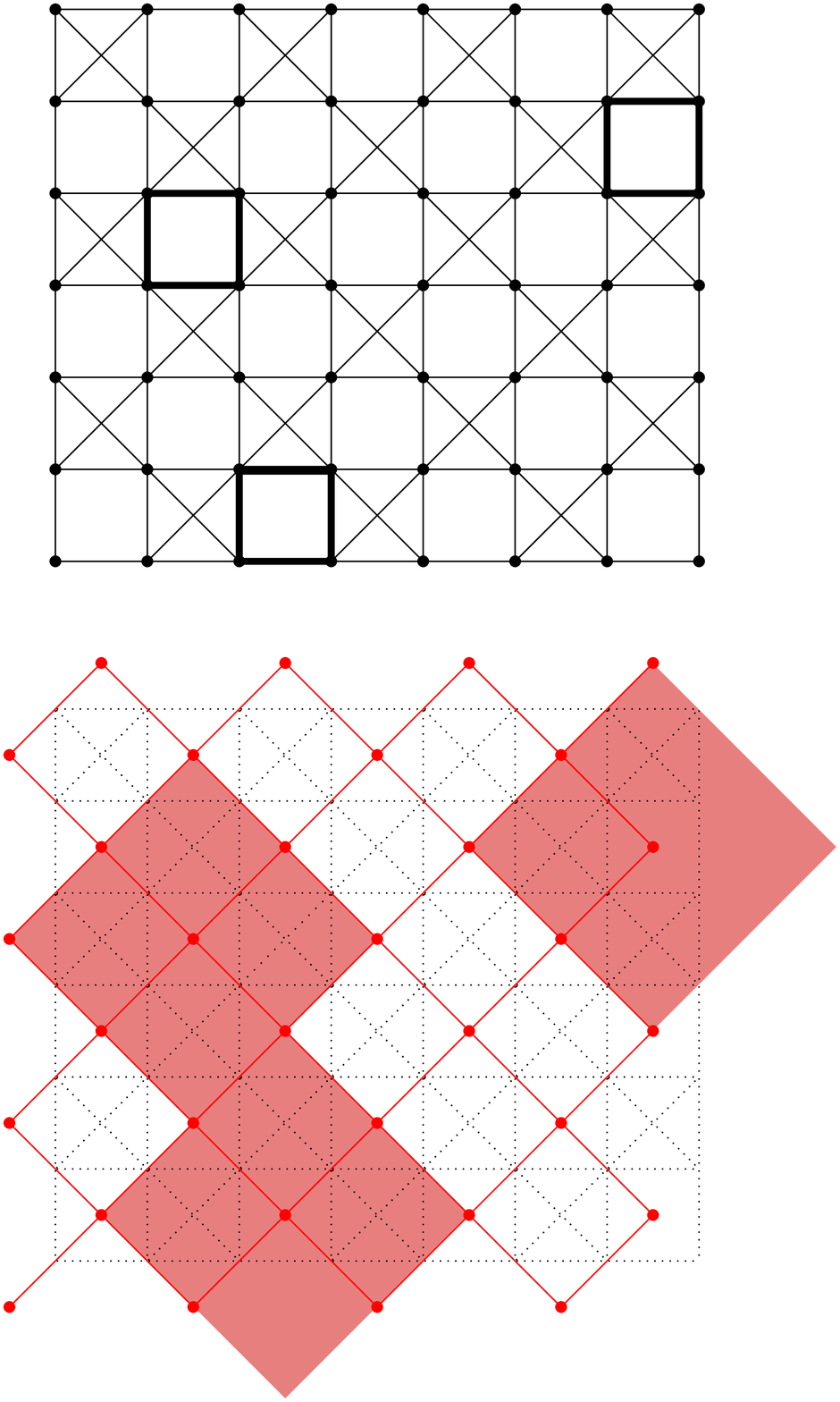}}
\caption
{The checkerboard lattice which hosts three localized magnons 
(bold squares)
and the auxiliary square lattice with hard squares 
used for the estimation of 
the ground-state degeneracy at saturation.
The localized magnons are eigenstates 
for exchange bonds of uniform strength  \cite{05}. 
\label{fig06}}
\end{figure}

To be specific,
we consider several geometrically frustrated lattices,
namely,
the dimer-plaquette chain \cite{14}
(Fig. \ref{fig01}),
\input epsf
the sawtooth chain \cite{15,16}
(Fig. \ref{fig02}),
\input epsf
two kagom\'{e}-like chains \cite{17,18}
(Figs. \ref{fig03}, \ref{fig04}),
\input epsf
\input epsf
the kagom\'{e} lattice
(Fig. \ref{fig05}),
\input epsf
and
the checkerboard (also called 2D or planar pyrochlore) lattice 
(Fig. \ref{fig06}).
\input epsf
The ground-state and low-temperature properties 
for the Heisenberg antiferromagnet on these lattices
are subjects of intensive discussions.
We consider $N$ quantum spins of length $s$
described by the Hamiltonian
\begin{eqnarray}
\label{01}
H=\sum_{(nm)}J_{nm}
\left(s_n^xs_m^x+s_n^ys_m^y+\Delta s_n^zs_m^z\right)
-h\sum_ns_n^z
\end{eqnarray} 
where
the first sum runs over the bonds (edges) which connect the sites (vertices)
occupied by spins for the mentioned lattices,
$J_{nm}>0$ is the antiferromagnetic exchange constant between neighboring sites,
$\Delta$ is the anisotropy parameter,
$h$ is the external magnetic field,
and the second sum runs over all sites.
We know from Refs. \onlinecite{04,05}, 
that for certain values of  
$J_{nm}$  the considered lattices host localized magnons (see also the
corresponding Figs. \ref{fig01} -- \ref{fig06}). 
Due to these localized magnons
the ground state of (\ref{01})
for the mentioned lattices at the saturation field $h_1$
is highly degenerate, since 
the energies of $n$ independent localized magnon states 
with $n=1,\ldots,n_{\max}$, $n_{\max}\sim N$
are exactly the same.
Because a certain local fragment of the lattice can be occupied by a 
magnon or not, 
the degeneracy of the ground state at saturation,
${\cal{W}}$,
grows exponentially with $N$ 
giving rise to a finite zero-temperature entropy per site at saturation
\begin{eqnarray}
\label{02}
\frac{{\cal{S}}}{k}=\lim_{N\to\infty}\frac{1}{N}\log{\cal{W}}.
\end{eqnarray}
The counting problem associated with the ground-state degeneracy
can be solved 
after mapping 
the lattice which hosts independent localized magnons 
onto some auxiliary lattice 
which is occupied by hard-core objects
(monomers, or monomers and dimers, or hexagons, or squares).

We start with the dimer-plaquette chain shown in Fig. \ref{fig01}.
The auxiliary lattice (Fig. \ref{fig01}, bottom) 
is a linear chain of ${\cal{N}}=\frac{1}{4}N$ sites 
which may be either occupied 
(if a localized magnon is trapped 
by the corresponding fragment of the initial lattice)
or empty
(in the opposite case).
Obviously,
\begin{eqnarray}
\label{03}
{\cal{W}}=2^{\cal{N}}
=\exp\left(\frac{1}{4}\log 2\;N\right)
\approx
\exp\left(0.173287 N\right).
\end{eqnarray}
Repeating these arguments for the diamond chain \cite{19} 
we arrive at the similar result, 
${\cal{W}}=2^{{\cal{N}}}$,
however,
with ${\cal{N}}=\frac{1}{3}N$.

Next we consider the sawtooth chain shown in Fig. \ref{fig02}.
The auxiliary chain (Fig. \ref{fig02}, bottom)
consists of ${\cal{N}}=\frac{1}{2}N$ sites 
which may be filled 
either by rigid monomers 
or by rigid dimers occupying two neighboring sites.
The limiting behavior of ${\cal{W}}$ for a large lattice 
${\cal{N}}\to\infty$
may be found in Ref. \onlinecite{07}
\begin{eqnarray}
\label{04}
{\cal{W}}
=\exp\left(\log\frac{1+\sqrt{5}}{2}\;{\cal{N}}\right)
\approx
\exp\left(0.240606 N\right).
\end{eqnarray}
The same result (\ref{04}) holds 
for the two-leg ladder of Refs. \onlinecite{20a,20b}
(see Fig.~1a of Ref. \onlinecite{20b}).
Similarly,
for the kagom\'{e}-like chains shown in Figs. \ref{fig03}, \ref{fig04}
we get
\begin{eqnarray}
\label{05}
{\cal{W}}
=\exp\left(\frac{1}{3}\log\frac{1+\sqrt{5}}{2}\;N\right)
\approx
\exp\left(0.160404 N\right)
\end{eqnarray}
and
\begin{eqnarray}
\label{06}
{\cal{W}}
=\exp\left(\frac{1}{5}\log\frac{1+\sqrt{5}}{2}\;N\right)
\approx
\exp\left(0.096242 N\right),
\end{eqnarray}
correspondingly.

Let us pass to the 2D case.
Considering the kagom\'{e} lattice (Fig. \ref{fig05}) 
we identify 
i) the centres of hexagons 
(which may trap magnons)
as the sites of the auxiliary triangular lattice 
and 
ii) the hexagons 
carrying localized magnons
together with the six attached triangles
as the shaded hexagons on the triangle lattice.
Now it is evident 
that the filling of the kagom\'{e} lattice by localized magnons 
corresponds 
to the occupation of the auxiliary triangular lattice by hard hexagons.
The hard-hexagon model 
(i.e. the triangular lattice gas with nearest-neighbor exclusion)
has been exactly solved \cite{09}.
In particular, 
for the number of ways of putting hard hexagons 
on the triangular lattice of ${\cal{N}}\to\infty$ sites
the accurate estimate is\cite{09}
$\exp\left(0.333 242 721 976 \ldots {\cal{N}}\right)$.
Therefore,
taking into account the relation 
between the number of sites $N$ of kagom\'{e} lattice 
and the number of sites ${\cal{N}}$ of the auxiliary triangular lattice,
$N=3{\cal{N}}$,
we get
\begin{eqnarray}
\label{07}
{\cal{W}}
\approx\exp\left(0.111081 N\right).
\end{eqnarray}
It should be noted here 
that the hard-hexagon model also arises 
while calculating ${\cal{W}}$ for the star lattice \cite{03,20},
however, in that case $N=6{\cal{N}}$.

Finally we consider the checkerboard lattice shown in Fig. \ref{fig06}.
The construction of an auxiliary lattice 
for the calculation of the ground-state degeneracy at saturation is illustrated 
in the lower part of Fig. \ref{fig06}.
Each centre of the square which may host a localized magnon 
is represented by a site of the auxiliary square lattice.
Moreover,
the square hosting a magnon together with the eight attached triangles 
of the checkerboard lattice
is represented by the shaded hard square 
(consisting of four elementary cells) 
of the auxiliary square lattice.
A one-to-one correspondence between independent localized  
magnon configurations 
and shaded hard-square configuration obviously exists.
As a result, 
we have to consider the square lattice gas with ${\cal{N}}=\frac{1}{2}N$
sites 
with nearest-neighbor and
next-nearest-neighbor exclusion.
We are not aware of an estimate of the entropy for such a model. 
(For the hard-square model 
(i.e. square lattice gas with
with only nearest-neighbor exclusion)
the number of ways of putting hard squares 
on the square lattice of ${\cal{N}}\to\infty$ sites
equals \cite{10}
$\exp\left(1.503 048 082 475 \ldots {\cal{N}}\right)$.)
A simple estimate for the lower bound for ${\cal{W}}$ is
$2^{\frac{N}{8}}\approx\exp\left(0.086643 N\right)$.

To summarize this part,
a class of frustrated quantum spin lattices 
has a huge degeneracy of the ground state at saturation 
that leads to a nonzero residual ground-state entropy.
For some of such models 
the zero-temperature entropy at saturation $h=h_1$
can be estimated exactly.
These values provide the ``reference points'' 
in the low-temperature dependence entropy ${\cal{S}}$ vs. field $h$
for the corresponding lattices.
It is remarkably 
that the calculation of ${\cal{W}}$ 
is a pure combinatorial problem 
and therefore the values of the ground-state entropy at saturation
are not sensitive 
to the value of anisotropy $\Delta$ or the value of spin $s$.

\begin{figure}[t]
\includegraphics[clip=on,width=12.cm,angle=-90]{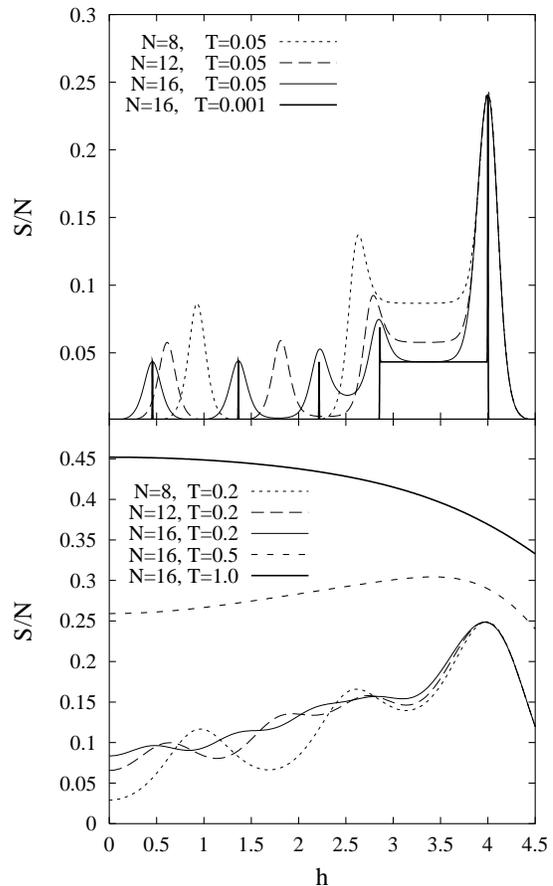}
\caption
{Field dependence of the isothermal entropy per site at low temperatures
(upper panel)
and higher temperatures
(lower panel)
for the sawtooth chains of different length
($s=\frac{1}{2}$, 
$\Delta=1$,
$J_1=1$, $J_2=2$).
\label{fig07}}
\end{figure}

\begin{figure}[t]
\includegraphics[clip=on,width=12.cm,angle=-90]{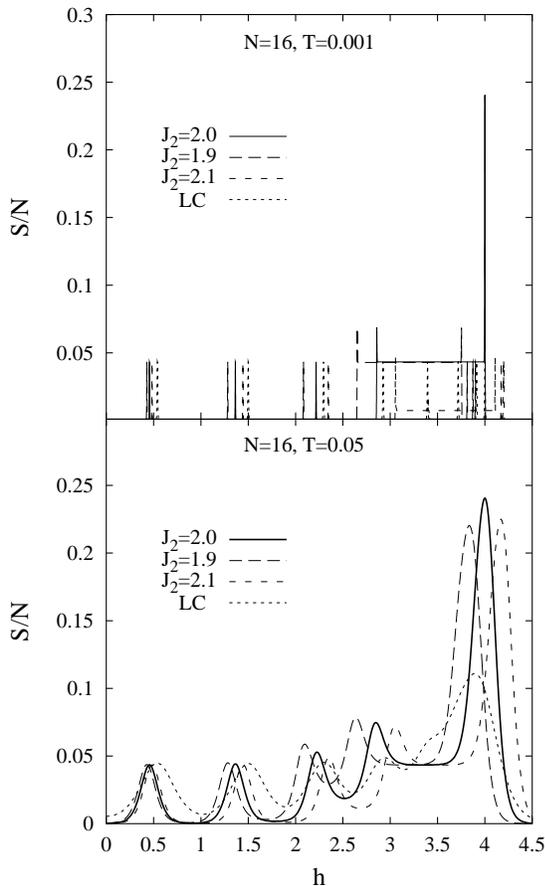}
\caption
{Field dependence of the isothermal
entropy per site of the sawtooth chain 
at very low temperature 
(upper panel)
and at higher (but still low) temperature
(lower panel) 
as $J_2$ deviates from $\sqrt{2\left(1+\Delta\right)}J_1$.
The corresponding dependence for a linear chain 
(LC, dotted curves)
is also reported 
for comparison.
\label{fig08}}
\end{figure}

In what follows we discuss the dependence of the entropy ${\cal{S}}$ on the
magnetic  field $h$ 
for  $h<h_1$ at arbitrary temperatures using full exact diagonalisation of
finite spin systems.
We expect that the qualitative behavior is similar for all lattices
considered. 
Here we focus on the sawtooth chain, because the ground-state 
degeneracy at saturation,
${\cal{W}}$, 
is largest and the finite-size effects should be smallest.
We have considered sawtooth chains
of $N=8,\;12,\;16$ sites
with $J_1=1$, $J_2=\sqrt{2\left(1+\Delta\right)}$,
anisotropy parameters $\Delta=1$ and $\Delta=0$,
spin lengths $s=\frac{1}{2},\;1,\;\frac{3}{2}$ 
at several temperatures 
$kT=0.001,\;0.05,\;0.2,\;0.5,\;1$.
Some of our numerical results 
are shown in Figs. \ref{fig07}, \ref{fig08}.

Let us discuss the obtained results. 
Firstly we note 
that for several magnetic fields below saturation $h<h_1$ 
one has a two-fold or even a three-fold degeneracy of the energy levels 
leading in a finite system to a finite zero-temperature  entropy.
Correspondingly  one finds in Fig. \ref{fig07} (upper panel)
a peaked structure and moreover a plateau just below saturation. However, it
is clearly seen in Fig. \ref{fig07} (upper panel)
that the height of the peaks and of the plateau decreases with system size $N$
and one has ${\cal{S}}=0$ at $T=0$ as $N\to\infty$ for $h<h_1$ and $h>h_1$, 
only the peak at $h=h_1$ does not vanish. 
At finite temperatures this peak survives as a well-pronounced 
maximum and it only disappears if the temperature grows up to 
the order of the exchange constant, Fig. \ref{fig07} (lower panel).
The value of entropy at saturation, 
which agrees with the analytical prediction (\ref{04}),
is almost temperature independent up to about $kT \approx 0.2$,
see Fig. \ref{fig07}.
Moreover,
the value of entropy at saturation is also almost size-independent
as it follows from data for different $N$.
Thus, 
the effect of the independent localized magnons 
which yield the residual ground-state entropy
survives at finite temperatures $kT \lesssim 0.2$
producing a noticeable enhancement in the isothermal entropy curve 
at the saturation field.

By calculating results for $\Delta=0$ and $\Delta=1$
and for $s=1,\;\frac{3}{2}$ and $s=\frac{1}{2}$
we have checked that the maximum  
in the entropy at saturation for low temperatures
is robust against 
exchange interaction anisotropy
and appears also  for larger spin values $s$.
However, our numerical results suggest 
that the enhancement of the entropy at saturation 
for finite temperatures becomes less pronounced with increasing $s$.
A simple reason for that could be the circumstance that the degeneracy at
saturation does not depend on spin value $s$, 
but the total number of state increases with $s$ according to $s^N$.

Concerning the  experimental confirmation 
of the predicted behavior of the entropy
in real compounds we are  faced with the situation that 
the conditions on bond strengths 
under which the independent localized magnons become the exact eigenstates
\cite{04,05,03}
are certainly
not strictly fulfilled.
For example,
for the isotropic Heisenberg sawtooth chain (\ref{01})
we have imposed $J_2=2J_1$, see Fig. \ref{fig07}.
Therefore,
it is useful to discuss  the ``stability'' of our conclusions 
against deviation from the perfect condition for bond strengths.
For this purpose we examine numerically
the field dependence of entropy at low temperatures
for the $s=\frac{1}{2}$ isotropic sawtooth chain of $N=16$ sites
with $J_1=1$ and $J_2=1.9$ and $J_2=2.1$ 
(Fig. \ref{fig08}).
Evidently,
the degeneracy of the ground state at saturation is lifted
when $J_2 \ne 2$
that immediately yields zero entropy at saturation at very low temperatures
(long-dashed and short-dashed curves 
in the upper panel of Fig. \ref{fig08}).
However,
the initially degenerate energy levels remain close to each other,
if $J_2$ only slightly deviates from the perfect value $2$. 
Therefore
with increasing temperature  
those levels become accessible for the spin system
and they manifest themselves in the entropy enhancement
in the vicinity of saturation
at low but nonzero temperatures. This
can be nicely seen 
in the lower panel in Fig. \ref{fig08}
(long-dashed and short-dashed peaks in the vicinity of saturation).
To demonstrate that this enhancement 
is the effect of the localized magnon states 
in the considered frustrated quantum spin lattice 
we also report the field-dependent entropy 
of the $s=\frac{1}{2}$ isotropic linear chain of $N=16$ sites
(dotted curves in Fig. \ref{fig08})
which remains in this field region 
at least two times smaller.

Let us remark
that the ground-state degeneracy problem 
of antiferromagnetic Ising lattices in the critical magnetic field 
(i.e. at the spin-flop transition point),
which obviously do not contain quantum fluctuations,  
has been discussed in the literature \cite{21}.
Thus, 
the exactly solvable case of the antiferromagnetic Ising chain 
at critical magnetic field 
provides another example 
of the low-temperature entropy enhancement at saturation.
From Ref. \onlinecite{21}
we know that at zero temperature
${\cal{S}}=k\log\frac{1+\sqrt{5}}{2}$ 
for $s=\frac{1}{2}$.
Repeating the transfer matrix calculations for $s=1$ and $s=\frac{3}{2}$ 
we find instead 
${\cal{S}}=k\log 2$
and
${\cal{S}}=k\log\frac{1+\sqrt{13}}{2}$,
respectively,
that shows 
that the zero-temperature entropy at critical field 
depends on the spin value $s$.
On the contrary, 
for the frustrated  quantum spin lattices 
considered in the present paper 
the zero-temperature entropy at saturation 
does not depend on $s$. 

Finally,
we should emphasize that there are other lattices 
which support the independent localized magnon states
e.g., the 2D square-kagom\'{e} lattice 
or the 3D pyrochlore lattice \cite{04,05}.
In these cases 
a rigorous result for the ground-state degeneracy at saturation
is not available, but for the existing 
huge degeneracy at saturation a lower bound is given by \cite{03} 
${\cal W} \ge 2^{n_{\max}}$ 
where $n_{\max}\sim N$
is the maximum number of localized magnons
which depends on the lattice geometry.   
This leads to the conclusion
that the discussed low-temperature peculiarity 
of the entropy in the vicinity of saturation 
should also be present. 
Hence, the low-temperature maximum  of $\cal{S}$
at saturation 
is a generic effect for strongly frustrated quantum 
spin lattices 
which may host independent localized magnons.

From the experimental point of view
the discussed effect of the independent localized magnons 
on the low-temperature field dependence of the entropy 
in the vicinity of saturation
may be of great importance.
Really,
although the most spectacular effect of the independent localized magnons 
is a jump in the zero-temperature magnetisation curve 
just below the saturation \cite{04,05,03},
it is probably difficult to observe the jump at finite temperatures.
The above discussed
maximum in the entropy vs. field curve due to
independent localized magnons
is certainly  easier accessible for experimental observation, since 
the isothermal entropy as a function of field 
can be obtained
from a specific-heat measurement  
(see, e.g. Refs. \onlinecite{12,13}). 
We also mention the significance of the  maximum in the entropy vs. field curve 
to an enhanced magnetocaloric effect \cite{22}.

To summarize,
we have rigorously calculated the finite ground-state entropy 
at the saturation field 
for some strongly frustrated quantum spin lattices 
hosting localized magnons.
To discuss the physical relevance of these results 
we have examined  
the field dependence of entropy 
at low temperatures for these  frustrated systems.
We have found that the independent localized magnon states 
produce a maximum in the isothermal  entropy versus field curve  
in the vicinity of the saturation field at low temperatures.
This effect is robust against small deviations 
from the condition on bond strengths 
under which the localized magnons exist.
The reported behavior can manifest itself 
in the high-field specific heat measurements 
permitting 
to detect experimentally the independent localized magnons 
in frustrated quantum spin lattices.

{\it Acknowledgments:}
We would like to thank R.~Moessner and A.~Honecker for fruitful discussions 
and J.~Schulenburg for assistance in numerical calculations.
We are in particular indebted to R.~Moessner
who brought our attention 
to the residual entropy calculation and the hard-hexagon problem.
J.~R. and O.~D. acknowledge the kind hospitality 
of the Max-Planck-Institut f\"{u}r Physik komplexer Systeme (Dresden)
in the spring of 2004.
This work was partly supported by the DFG (project Ri615/12-1).

\end{document}